\documentstyle[epsfig,12pt]{article}

\newcommand{\bce}{\begin{center}} 
\newcommand{\ece}{\end{center}}
\newcommand{\beq}{\begin{equation}}
\newcommand{\eeq}{\end{equation}}
\newcommand{\bea}{\vspace{0.25cm}\begin{eqnarray}}
\newcommand{\eea}{\end{eqnarray}}

\newcommand{\ba}{\begin{array}}
\newcommand{\ea}{\end{array}}

\newcommand{\doublespace}{
    \renewcommand{\baselinestretch}{1.6}\large\normalsize}

\def\lsim{\mathrel{\rlap{\lower4pt\hbox{\hskip1pt$\sim$}}
    \raise1pt\hbox{$<$}}}         
\def\gsim{\mathrel{\rlap{\lower4pt\hbox{\hskip1pt$\sim$}}
    \raise1pt\hbox{$>$}}}         

\def\Pom{{\bf I\!P}}

\def\beq{\begin{equation}}
\def\endeq{\end{equation}}
\def\arr{\begin{eqnarray}}
\def\endarr{\end{eqnarray}}
\makeindex

  \advance\oddsidemargin  by -1in
  \advance\evensidemargin by -1in
  \advance\topmargin      by -0.5in
\begin{document}

\vspace{2.0cm}

\begin{flushright}
\end{flushright}

\vspace{1.0cm}

\begin{center}
{\Large \bf  The BFKL-Regge factorization and $F_2^b$, $F_2^c$, $F_L$ at 
HERA:\\  physics implications of nodal properties of the BFKL eigenfunctions}

\vspace{1.0cm}

{\large\bf R.~Fiore$^{1 \dagger}$, N.N. Nikolaev$^{2 \ddagger}$ and 
V.R.~Zoller$^{3 \times}$}

\vspace{1.0cm}

$^{1)}${\em Dipartimento di Fisica,
Universit\`a     della Calabria\\
and\\
 Istituto Nazionale
di Fisica Nucleare, Gruppo collegato di Cosenza,\\
I-87036 Rende, Cosenza, Italy}\\

$^{2)}${ \em
Institut  f\"ur Kernphysik, Forschungszentrum J\"ulich,
D-52425 J\"ulich, Germany\\
and\\
L.D.Landau Institute for Theoretical Physics, Chernogolovka,
142432 Moscow Region, Russia}\\

$^{3)}${\em ITEP, Moscow 117218, Russia}

\vspace{1.0cm}

{\bf Abstract}
\end{center}
The asymptotic freedom is known to split  the leading-$\log$ BFKL pomeron
 into a  series of isolated poles in the complex angular  momentum plane.
One of our earlier findings was that the subleading hard BFKL exchanges 
decouple from such experimentally important observables as small-$x$
charm, $F_2^c$,  and the longitudinal, $F_L$, structure functions  of the 
proton at moderately large $Q^2$. For instance, we predicted precocious 
BFKL asymptotics of $F_2^c(x,Q^2)$ with intercept of the rightmost  BFKL pole 
$\alpha_{\Pom}(0)-1=\Delta_{\Pom}\approx 0.4$. On the other hand, the
small-$x$ open beauty photo- and electro-production probes the vacuum 
exchange for much smaller color dipoles which entails significant  
subleading  vacuum pole corrections to the small-$x$ behavior. 
In view of the accumulation of the 
experimental data on small-$x$ $F_{2}^{c}$, $F_{2}^{b}$ and $F_{L}$  we extend 
our early  predictions to the kinematical domain covered by new HERA measurements.
Our parameter-free results  agree well  with the  determination 
of  $F_2^c$, $F_L$  and published H1 results on $F_2^b$  
but slightly overshoot the very recent (2008, preliminary) 
H1 results on $F_2^b$.

\doublespace

\vskip 0.5cm \vfill $\begin{array}{ll}
^{\dagger}\mbox{{\it email address:}} & \mbox{fiore@cs.infn.it} \\
^{\dagger}\mbox{{\it email address:}} & \mbox{n.nikolaev@fz-juelich.de} \\
^{\times}\mbox{{\it email address:}} & \mbox{zoller@itep.ru} \\
\end{array}$

\pagebreak


{\bf 1.}  Within the color-dipole (CD) approach to the BFKL pomeron, the flavor
independence is a fundamental feature of the dipole cross section, 
while the QCD pomeron contribution
would depend on the interacting particles through the QCD impact factors, calculable in
terms of the flavor-dependent color dipole structure of the target and projectile.
As noticed by Fadin, 
Kuraev and Lipatov in 1975 (\cite{FKL}, see also more detailed discussion by
Lipatov \cite{Lipatov}), incorporation of the asymptotic freedom into
the QCD BFKL equation splits the fixed-$\alpha_S$ cut in the complex $j$-plane
into a series of
isolated BFKL-Regge poles. 
Such a spectrum has a far-reaching theoretical
and experimental consequences because
a contribution of each isolated hard BFKL pole to the scattering amplitudes
and/or  structure functions (SF) would satisfy a very powerful 
Regge factorization \cite{Gribov}.
The resulting CD BFKL-Regge factorized expansion allows one to relate in a
parameter-free fashion SF's of different targets, 
$p,\pi,\gamma,\gamma^{*}$ \cite{DER,PION,GamGam} and/or 
contributions of different flavors to 
the proton SF \cite{CHARM2000, BEAUTY}. Within  the color
dipole formulation of the BFKL equation \cite{BFKL} the first analysis of
small-$x$ behavior of
open charm SF of the proton, $F_2^c$,
 in the color
dipole formulation of the BFKL equation \cite{BFKL} has been carried out in 1994 
\cite{NZZ94,NZDelta,NZHERA} with an intriguing result that for moderately
 large 
$Q^{2}$ it is dominated by the leading hard BFKL pole  exchange.
Later on this fundamental 
feature of CD
BFKL approach has been related \cite{NZcharm} to nodal properties 
of eigen-functions of
subleading hard BFKL-Regge poles \cite{NZZ97}.

In \cite{NZZ97}  the latter
property of the CD BFKL-Regge factorization was applied
   and the strength of
the subleading hard BFKL corrections  and soft-pomeron background to 
dominant rightmost hard BFKL exchange was quantified. 
One of the observations  of Ref.~\cite{NZZ97} 
is that the 
BFKL-Regge expansion (\ref{eq:3.3}) truncated at $m=2$ appears to be very successful
 in describing of the proton SF's in a wide range of $Q^2$. Very recently this
phenomenon  has been rediscovered in Ref.~\cite{ELLIS}.

In view of the  accumulation of the experimental data on small-$x$
$F_{2}^{c}$, $F_{2}^{b}$   we extended 
 early predictions to 
the kinematical domain covered by new HERA measurements.
 Based on the CD
BFKL-Regge factorization we report a parameter-free description of  both
$F_2^c$ and $F_2^b$. A specific feature of our CD approach is a 
 decoupling of soft and
subleading BFKL singularities at the scale of the open charm production
 which entails a precocious asymptotic 
 BFKL behavior of the
the structure function $F_2^c$. Reversing the argument, the 
open charm excitation by real photons and in DIS 
gives a particularly clean access to the intercept of the rightmost hard 
BFKL pole \cite{NZZ94}. Here we  show how  the
interplay of leading and subleading vacuum exchanges predicts a rise of the
beauty structure function of the proton $F_2^b$ much faster than 
prescribed by the leading pomeron trajectory 
 (see also the early discussion in Ref.~\cite{BEAUTY}). All our predictions 
are parameter-free and we find a nice agreement with the 
published experimental data from H1
Collaboration \cite{H1ccbb} on the charm and beauty
SF of the proton, although the  very recent 
preliminary  H1 results on  $F_2^b$
\cite{H12008} are slightly over-predicted. .
The  longitudinal structure function of the proton $F_L$ is still another 
observable selective of the dipole size and we report 
the BFKL-Regge factorization results for $F_L$.
 The recent H1 measurements of $F_L$ \cite{H1FL2008} 
are consistent with our 
predictions made in \cite{CHARM2000} but  are too uncertain for 
any firm conclusions. Taken together, the experimental data on hard 
structure functions do strongly corroborate our 
1994 prediction 
$\Delta_{\Pom}\approx 0.4$ for the intercept of the 
rightmost hard BFKL pole.



{\bf 2.} Within the CD approach to small-$x$ DIS excitation of heavy flavor 
is described by interactions of $q\bar{q}$ color
dipoles in the photon of  a predominantly small size  ${\bf r}$,
\beq
{ 4 \over Q^{2}+4m_{q}^{2}}\lsim r^{2} \lsim { 1\over m_{q}^{2}}\, ,
\label{eq:1.1}
\eeq
which makes them an arguably sensitive probe of the 
short distance properties of the vacuum exchange in QCD in the 
Regge regime 
\beq
{1\over x}={W^2+Q^2\over4 m^2_c + Q^2}\gg 1\,.
\label{eq:1.2}
\eeq
The CD cross section $\sigma(x,{\bf r})$ depends neither on flavor, nor beam,
nor target, and the contribution of excitation of the open 
charm/beauty to photo-absorption cross section is given by the color dipole
factorization formula
\beq
\sigma^{c}(x,Q^{2})=
\int dz d^{2}{\bf{r}} 
|\Psi_{\gamma^*}^{~c\bar{c}}(z,{\bf{r}})|^{2} 
\sigma(x,{\bf{r}})\,.
\label{eq:2.1}
\eeq
\begin{figure}[h]
\psfig{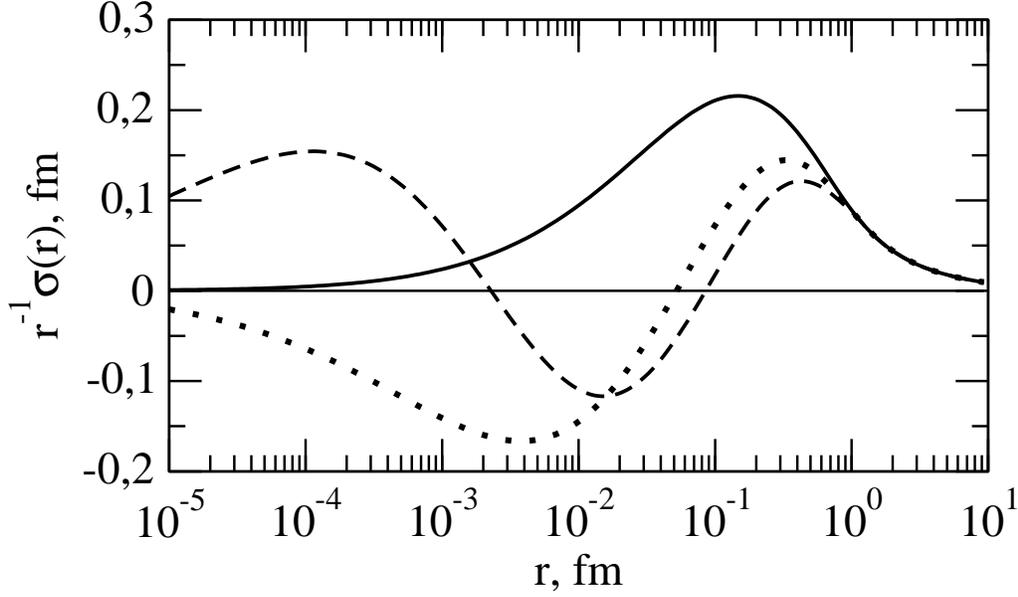}
\vspace{-0.5cm}
\caption{The rightmost $\sigma_0/r$ and subleading $\sigma_1/r$ and  
$\sigma_2/r$
eigen cross sections as a function of $r$} are shown. 
\label{fig:fig1}
\end{figure} 
 
Here $|\Psi_{\gamma^*}^{~c\bar{c}}(z,{\bf{r}})|^{2}$  is
the probability to find in the photon the $c\bar{c}$ color dipole with the
charmed quark carrying the fraction $z$ ofthe light-cone momentum of the photo
 \cite{NZ91}.  Hereafter we focus on the charm structure
function  
\bea
F_2^{c}(x_{Bj},Q^2)=
{Q^2\over {4\pi^2\alpha_{em}}}\sigma^{c}(x,Q^{2})
=\int{ d r^{2}\over r^{2}}{\sigma(x,r) \over r^{2}} W_{2}(Q^{2},m_{c}^{2},r^2)
\,.
\label{eq:2.4}
\eea
 A detailed  analysis of  the weight function $W_{2}(Q^{2},m_{c}^{2},r^2)$ is
found in Refs.~\cite{NZDelta,NZHERA}, we only cite
the principal results: (i) at moderate $Q^{2} \lsim 4m_{c}^{2}$ the weight
function has a peak at a scanning radius $r =r_S \sim {1/ m_{c}}$, (ii) at very high $Q^{2}$ the 
peak develops a plateau for dipole sizes in the interval (\ref{eq:1.1}), (iii)
the contribution from large dipoles is strongly suppressed 
for heavy flavors.
One can say that for moderately large $Q^{2}$ excitation of open charm 
probes (scans) the dipole cross section at a special dipole size $r_{S}$
(the scanning radius) 
\beq
r_{S} \sim {1/ m_{c}}\,.
\label{eq:2.6}
\eeq



{\bf 3.} In the Regge region of ${1\over x} \gg 1$ the CD cross section $\sigma(x,r)$
satisfies the CD BFKL equation
\beq
{\partial \sigma(x,r) \over \partial \log{(1/x)}}={\cal K}\otimes \sigma(x,r)\;
\label{eq:3.1}
\eeq
for the kernel ${\cal K}$ of CD approach see Ref.~\cite{PISMA1}.
The solutions with Regge behavior
\beq
\sigma_{m}(x,r)=
\sigma_{m}(r)\left({1\over x}\right)^{\Delta_{m}}
\label{eq:3.0} 
\eeq
satisfy the eigen-value 
problem
${\cal K}\otimes \sigma_{m}=\Delta_{m}\sigma_{m}(r)$
and the CD BFKL-Regge expansion for the 
color dipole cross section
reads \cite{NZZ94,PION}
\beq
\sigma(x, r)=\sum_{m=0} 
\sigma_m(r)\left({x_0\over x}\right)^{\Delta_m}.
\label{eq:3.3}
\eeq

 The practical calculation of $\sigma(x,r)$
    requires the boundary condition $\sigma(x_0,r)$ 
at certain $x_0\ll 1$.
We take for boundary condition at $x=x_0$ the Born approximation,
$\sigma(x_0,r)=\sigma_{Born}(r)\,,$ 
 i.e. evaluate dipole-proton scattering via the two-gluon exchange.
This leaves the starting point $x_0$ the sole parameter. 
The choice $x_0=0.03$ met a remarkable
 phenomenological success 
\cite{NZZ97,DER,PION}.

The properties of our CD BFKL equation and the choice of
physics motivated boundary condition were discussed in detail elsewhere 
\cite{NZDelta,NZHERA,NZcharm,NZZ97,DER}, 
here we only recapitulate features 
relevant to the considered problem. Incorporation of asymptotic freedom 
exacerbates the well known infrared sensitivity of
the BFKL equation and infrared regularization by infrared freezing of the 
running coupling $\alpha_S(r)$ and modeling of confinement of gluons 
by the finite propagation radius of perturbative gluons $R_c$ need to be
 invoked.

The leading eigen-function 
$\sigma_0(r)\equiv\sigma_{\Pom}(r)$
for ground state i.e., for the rightmost hard BFKL pole is node free.  
The  subleading  eigen-function for the excited state $\sigma_m(r)$ has $m$ nodes
(see Fig.~\ref{fig:fig1}).
We solve for $\sigma_m(r)$ numerically \cite{NZZ97,DER} (for the semi-classical 
analysis see the paper of Lipatov of Ref.~\cite{Lipatov}.
The so found intercepts (binding energies) follow
to a good approximation the law of Lipatov, 
$\Delta_{m}= \Delta_{0}/(m+1).$
 For the 
preferred 
$R_c=0.27\, {\rm fm}$ as chosen in 1994 in Refs.~\cite{NZHERA,NZDelta} and 
supported by
the  analysis \cite{MEGGI} of lattice QCD data 
we find 
$\Delta_{0}\equiv\Delta_{\Pom}=0.4\,.$  
The  node of $\sigma_{1}(r)$ is located at $r=r_1\simeq 
0.056\,{\rm fm}$, for larger $m$ the rightmost node moves to a somewhat  
larger $r=r_1\sim 0.1\, {\rm fm}$. The second node of eigen-functions with
$m= 2,3$ is located at  $r_{2}\sim 3\cdot 10^{-3}~ {\rm fm}$ which corresponds
to the momentum transfer scale 
$Q^{2} ={1/r_{2}^{2}}=5\cdot 10^{3}$ GeV$^{2}$.
The third node of $\sigma_{3}(r)$ is located at $r$ beyond the reach of any
feasible DIS experiments. It has been found \cite{NZZ97} that the BFKL-Regge
 expansion (\ref{eq:3.3}) truncated at $m=2$ appears to be very successful
 in describing of the proton SF's at 
  $Q^2\lsim 200$ GeV$^2$. However, at higher $Q^2$ and
moderately
small $x\sim x_0=0.03$ the background of the CD BFKL solutions with
 smaller intercepts ($\Delta_m < 0.1$)
should be taken into account (see below). 

Now comes the crucial observation that numerically 
$r_{1} \sim  r_{S}$. 
 Consequently, in the calculation of the open 
charm eigen-SF's  one scans the eigen-cross 
section in the vicinity of the node, which leads to a strong suppression of
the 
subleading contributions. 
\begin{figure}[h]
\psfig{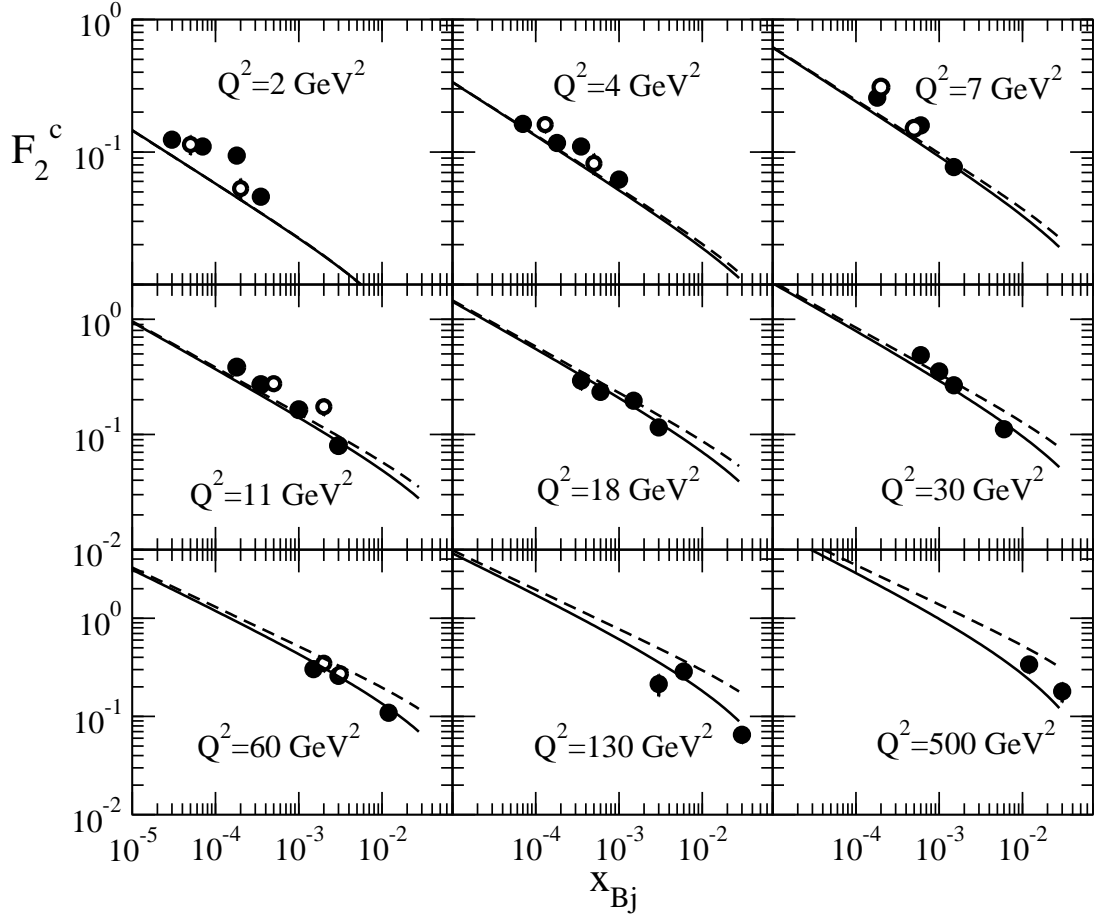}
\vspace{-0.5cm}
\caption{The prediction from CD BFKL-Regge factorization for the  charm structure
 function 
of the proton $F^{c}_2(x,Q^2)$ as a function of
the Bjorken variable $x_{Bj}$ is compared with the experimental data from 
 H1  Collaboration \cite{H1ccbb}. The  solid curve represents the result of the complete
CD BFKL-Regge expansion, the contribution of the rightmost hard BFKL pole  
 with $\Delta_{\Pom}=0.4$ is shown by the dashed line}
\label{fig:fig2}
\end{figure}  


{\bf 4.} Because the probability to find large color dipoles in the photon decreases
rapidly with the quark mass, the contribution from soft-pomeron exchange
to open charm excitation is very small down to $Q^{2}=0$.  
As we discussed elsewhere \cite{PION,CHARM2000}, for still higher solutions, 
$m\geq 3$, 
all intercepts are very small anyway, $\Delta_m\ll \Delta_{0}$,
For
this reason, on the purpose of practical phenomenology, we  truncate the 
expansion (\ref{eq:3.3}) at $m=3$ lumping in the term $m=3$ the contributions 
of still higher singularities with $m\geq 3$. The term $m=3$
is  endowed
 with the effective intercept $\Delta_3=0.06$
 and is presented in Ref.~ \cite{CHARM2000}
 in its analytical form.

We comment first on the results on $F_2^c$. The solid curve in Fig. \ref{fig:fig2}
is the result of the complete CD BFKL-Regge expansion. The dashed 
curve shows the pure rightmost hard BFKL pomeron contribution, in the Leading Hard
Approximation (LHA). 
There is a strong cancellation between soft and  subleading 
contributions with $m=1$ and $m=3$. Consequently, for this dynamical reason 
in this region of $Q^{2} \lsim 10$ GeV$^{2}$ we have an effective one-pole 
picture and LHA gives reasonable description of  $F_2^c$.

  In agreement with the nodal structure of subleading
eigen-SF's discussed in Refs.~\cite{PION,CHARM2000}, the LHA over-predicts slightly 
 $F_2^c$ at $Q^{2} \gsim 
30$ GeV$^{2}$. Here the negative valued subleading hard BFKL exchanges
overtake the soft-pomeron exchange  and the background from 
subleading hard BFKL exchanges becomes substantial at $Q^{2} \gsim 30$ GeV$^2$
and would even dominate $F_2^c$ at $Q^{2}\gsim 200$
GeV$^{2}$ and $x\gsim 10^{-2}$. In this region of $Q^{2}$ the soft-pomeron
exchange is numerically negligible.
The aforementioned
soft-subleading cancellations at $Q^{2} \lsim 
20$ GeV$^{2}$ become less accurate at smaller $x$, but here 
 both soft and subleading hard BFKL exchanges  become 
 Regge suppressed, because proportional to $x^{\Delta_{\Pom}}, 
~x^{\Delta_{\Pom}/2}$, respectively. 

In Fig.~\ref{fig:fig2} we compare our CD BFKL-Regge predictions to the 
recent experimental data from the H1 Collaboration \cite{H1ccbb} and 
find a very good agreement between the theory and experiment which
lends support to our 1994 evaluation $\Delta_{\Pom}=0.4$. The negative 
valued contribution from the subleading hard BFKL exchange is important for
bringing the theory to agreement with the experiment at large $Q^{2}$.
 For an alternative interpretation of heavy flavor production
see Refs.~\cite{TOLYA, White, Baranov} and references therein.
\begin{figure}[h]
\psfig{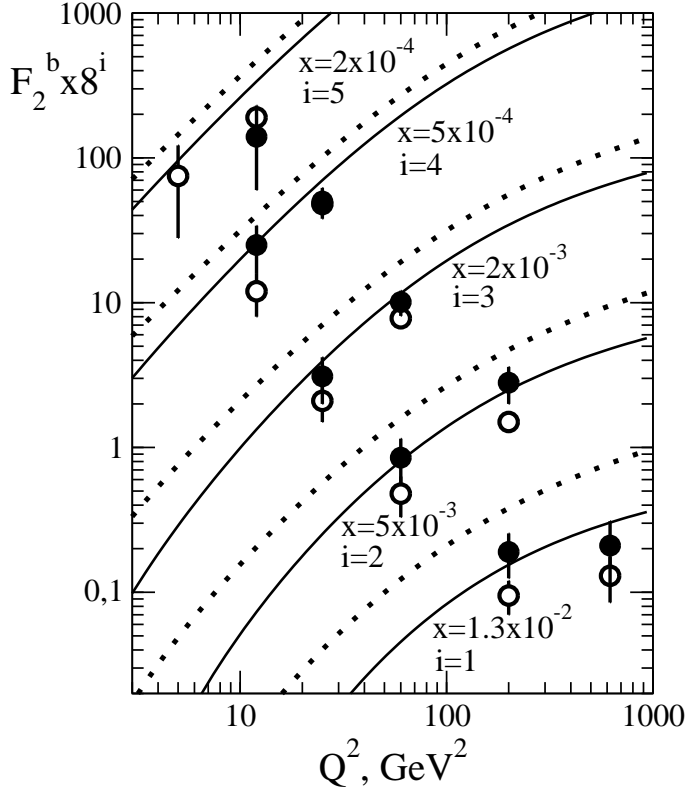}
\vspace{-0.5cm}
\caption{The predictions from CD BFKL-Regge factorization 
for the beauty 
structure function $F^{b}_2(x,Q^2)$, compared with data o
Refs.~\cite{H1ccbb}
(full circles) and 
\cite{H12008} (open circles) are shown. The  solid curve is the result of the complete
CD BFKL-Regge expansion, the contribution of the rightmost hard BFKL pole  
with $\Delta_{\Pom}=0.4$ is repesented by dotted curves.} 
\label{fig:fig3}
\end{figure} 

{\bf 5.} The characteristic feature of the QCD pomeron
 dynamics at distances $\sim m_b^{-1}$ is the large negative valued 
 contribution to
$F_2^b$,  coming from subleading BFKL singularities, 
see Fig. \ref{fig:fig1} and Ref.~\cite{BEAUTY}. 
 Consequences of this observation
 for the exponent of the
 energy dependence of the structure function 
\beq
F_2^b\propto \left(x_0\over x\right)^{\Delta_{\rm eff}}
\label{eq:SBBEXP}
\eeq
 are quite interesting.
In terms of the ratio
$r_m=\sigma_m/\sigma_0$ (see Fig. {\ref{fig:fig1})
 the exponent
$\Delta_{\rm eff}$ reads (m=1,2,3,soft)\cite{BEAUTY}
\beq
\Delta_{\rm eff}=\Delta_{0}\left[1-\sum_{m=1}r_m(1-\Delta_m/\Delta_0)
({x_0/ x})^{\Delta_m-\Delta_0}\right]
\label{eq:DELTAEFF}
\eeq
 Coefficients $r_m$
 in Eq.~(\ref{eq:DELTAEFF}) depend on $r$. They are negative on the left from
the rightmost node (Fig. \ref{fig:fig1}) and positive on the right.
Because for  $r\sim m_b^{-1}$ 
all $r_m$
are negative, except $r_{\rm soft}(0)>0$ \cite{BEAUTY}, at HERA energies
 the effective
 intercept $\Delta_{\rm eff}\equiv \Delta_{\rm beauty}$ 
overshoots the asymptotic value
 $\Delta_{\Pom}\equiv \Delta_0=0.4\,.$
 At still higher collision energies  both  the soft and subleading hard BFKL
  exchanges become rapidly Regge suppressed and we expect 
 $\Delta_{\rm eff}$ to decrease down to
 $\Delta_{\Pom}$ \cite{BEAUTY}.
This must be constructed to an aforementioned positive valued 
 subleading BFKL and  soft terms in the CD BFKL-Regge expansion
for light flavor SF's of the proton (see \cite{PION} for more details), which
lowers the pre-asymptotic
 pomeron intercept in  photoproduction of light flavors.
Hence the CD prediction of the  hierarchy of  pre-asymptotic 
intercepts \cite{BEAUTY} is 
\beq
\Delta_{\rm beauty}>\Delta_{\rm charm}>\Delta_{\rm light}\, .
\label{eq:BCL}
\eeq 
\begin{figure}[h]
\psfig{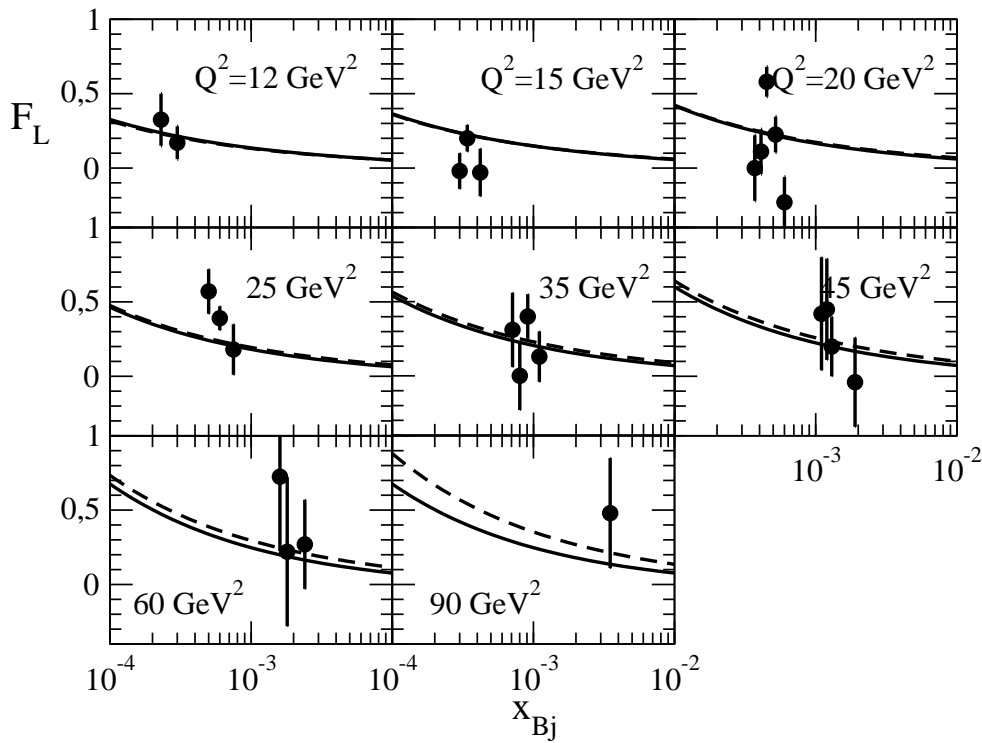}
\caption{The prediction from CD BFKL-Regge factorization for the longitudinal 
structure function 
of the proton $F_L(x,Q^2)$ as a function of
the Bjorken variable $x_{Bj}$ is shown. The  solid curve is the result of the complete
CD BFKL-Regge expansion, the contribution of the rightmost hard BFKL pole  
with $\Delta_{\Pom}=0.4$ is represented by dashed line. Data points ar
from Ref.~\cite{H1FL2008}} 
\label{fig:fig4}
\end{figure}
In Fig. \ref{fig:fig3}  we presented our predictions for the beauty 
structure function.
 The solid curve corresponds to the
complete expansion (\ref{eq:3.3}) while 
 the dotted
curve is the LHA.
 In agreement with the nodal structure of subleading
eigen-SF's the latter  over-predicts
 $F_2^b$ significantly because the negative valued contribution from
 subleading hard BFKL
 exchanges
overtakes the soft-pomeron exchange and the background from
subleading hard BFKL exchanges is substantial for all $Q^{2}$ values
 \cite{BEAUTY}.
Our    predictions for the beauty SF
 agree well  with the determination of  $F_2^b$  
by the H1 collaboration published in 2006 \cite{H1ccbb} (full circles)
 but slightly
overshoot smaller values of  $F_2^b$ from the  very 
recent  preliminary  
H1 results on $F_2^b$ reported in 2008 \cite{H12008} (open circles).
  

{\bf 6.} The cross section of diffractive (elastic) $\Upsilon(1S)$ meson
photoproduction has been measured at HERA \cite{ELBEAUTY}. Quarks in
 $\Upsilon$ meson are nonrelativistic and for real photons the $b\bar{b}$ CD 
wave function of the photon  $|\gamma\rangle is proportional to 
 m_bK_0(m_br)$, where $K_0(x)$ is the Bessel-MacDonald 
function \cite{SCAN}. The forward $\gamma\to \Upsilon$ transition matrix element
 $\langle\Upsilon|\sigma_n(r)|\gamma\rangle$ is controlled by the
 product $\sigma_0(r)K_0(m_br)$ \cite{SCAN}
and the amplitude of  elastic
  $\Upsilon(1S)$ photoproduction
  is dominated by the contribution from the dipole
sizes $r\sim r_{\Upsilon}=A/m_{\Upsilon}$
with $A \approx 5$. For a recent review on diffractive vector mesons see
Ref.~\cite{VMreview}. 

The crucial observation   is that at distances 
$r\sim r_{\Upsilon}$
  cancellations between soft and subleading contributions to the elastic
photoproduction cross section  result in the exponent
  $\Delta$ in 
\beq
{d\sigma(\gamma p\to\Upsilon p )\over dt}|_{t=0}\propto W^{4\Delta}
\label{eq:ELBB}
\eeq
 which is very close to $\Delta_{\Pom}$,  $\Delta=0.38$ \cite{BEAUTY,JETPVM}. This
observation  appears  to be in
agreement with the cross section rise observed by ZEUS and H
collaborations 
\cite{ELBEAUTY}.

{\bf 7.} It has been demonstrated in Ref.~\cite{NZDelta} that the longitudinal structure
function $F_L(x,Q^2)$ emerges as local probe of the dipole
cross section at $r^2\simeq 11/Q^2$.
The subleading CD BFKL cross sections have their rightmost node
at $r_1\sim 0.05-0.1$ fm. Therefore, one can zoom at the leading CD BFKL  pole
contribution and measure the pomeron intercept
$\Delta_{\Pom}$ from the $x$-dependence of $F_L(x,Q^2)$
at $Q^2\sim 10-30$ GeV$^2$. 
The aforementioned soft-subleading cancellation
 is nearly exact at $Q^2\sim 10-30$ GeV$^2$ and we predict a leading 
hard pole dominance in this region, as one can see from 
see Fig. \ref{fig:fig4}), where comparison with the very recent 
H1 data \cite{H1FL2008} is presented. We predicted the correct magnitude of 
 $F_L(x,Q^2)$, although the experimental data do not allow to draw 
conclusions on the $x$-dependence.


{\bf 8.} A simple note in passing. Okun and Pomeranchuk argued that for the members of
the same isotopic multiplet the strong interaction cross sections would have
an identical high energy behavior \cite{OP}.
Compare  the total cross section of the  charged and neutral components of
isotriplets of mesons like the $\rho$-mesons or pions  on an electrically
neutral target like a neutron. Arguably, for such a target the electromagnetic
breaking of the Okun-Pomeranchuk theorem  will be dominated by the
electromagnetic lifting of the  degeneracy of sizes of the charged and neutral
$\rho$'s. The strength of the Coulomb interaction in the charge and neutral
mesons is proportional to $e_u e_d$ and $-(e_u^2 +e_d^2)/2$,  respectively, the
net difference being proportional to $(e_u+e_d)^2$. Consequently, the difference of the
radii mean squared can be estimated as $\sim  \alpha_{em}(e_u+e_d)^2\langle r^2
\rangle$, what would entail
\beq
{{\sigma_{\pm} - \sigma_o } \over {\sigma_{\pm} + \sigma_o }} \sim 
\alpha_{em}(e_u+e_d)^2~.
\label{eq:OKUN1}
\eeq


{\bf 9.} The color dipole approach to the BFKL dynamics predicts uniquely a 
decoupling 
of subleading hard BFKL exchanges from open charm SF of
the proton at $Q^2\lsim 20\,{\rm GeV^2}$ and from $F_L$ at 
$Q^2\simeq 20\,{\rm GeV^2}$. This decoupling is due to a
dynamical cancellation between contributions of different subleading
hard BFKL poles and leaves us with an effective soft+rightmost hard BFKL 
two-pole approximation with intercept of the 
soft pomeron $\Delta_{{\rm soft}}=0$. 
We predict strong cancellation between the soft-pomeron 
and subleading hard BFKL contribution to $F_2^c$ in the experimentally
 interesting region
of $ Q^{2} \lsim 20$ GeV$^{2}$, in which $F_2^c$
is dominated entirely by the contribution from the rightmost hard BFKL pole.
This makes open charm in DIS at $Q^2\lsim 20$ GeV$^{2}$ a unique 
handle on the intercept of the rightmost hard BFKL exchange.
Similar hard BFKL pole dominance holds for $F_L(x,Q^2)$ .

High-energy open beauty
photoproduction probes the behavior $r \sim 1/m_b$ for the color dipole size 
and picks up a significant  contribution from the  subleading BFKL
poles. This makes  $\sigma^{b\bar b}(W)$ to grow much faster than it is
prescribed by the leading BFKL pole with an
intercept $\alpha_{\Pom}(0)-1=\Delta_{\Pom}=0.4$. Our calculations
within the  color dipole BFKL model are in
agreement with the recent determination of $\sigma^{b\bar b}(W)$ by the H1
collaboration.  The comparative analysis
 of diffractive photoproduction of  beauty, charm and light quarks exhibits
the hierarchy of pre-asymptotic pomeron intercepts  which follows the
hierarchy of corresponding hardness scales.
  We comment  on the phenomenon of decoupling of soft and
subleading BFKL singularities at the scale of elastic $\Upsilon(1S)$
 -photoproduction which results in precocious color dipole
 BFKL asymptotics of the
process $\gamma p \to \Upsilon p$.
 The agreement with the 
presently available
experimental data on open charm/beauty  in DIS confirm
 the CD BFKL prediction of the intercept $\Delta_{\Pom}=0.4$ 
for the rightmost hard BFKL-Regge pole.\\

{\bf Acknowledgments: } 
V.R.~Z. thanks  the Dipartimento di Fisica dell'Universit\`a
della Calabria and the Istituto Nazionale di Fisica
Nucleare - gruppo collegato di Cosenza for their warm
hospitality while a part of this work was done.
The work was supported in part by the Ministero Italiano
dell'Istruzione, dell'Universit\`a e della Ricerca and  by
 the RFBR grants 07-02-00021 and 09-02-00732.





\begin{thebibliography}{99}
\bibitem{FKL}
 V.S. Fadin, E.A. Kuraev, and L.N. Lipatov  {\sl Phys. Lett.}
B{\bf 60}, 50 (1975);
E.A. Kuraev, L.N. Lipatov, and V.S. Fadin, {\sl Sov. Phys. JETP}
{\bf 44}, 443 (1976); {\bf 45}, 199 (1977).

\bibitem{Lipatov}
L.N. Lipatov, {\sl Sov. Phys. JETP} {\bf 63}, 904 (1986).

\bibitem{Gribov}
V.N. Gribov, B.L. Ioffe, I.Ya. Pomeranchuk, and A.P. Rudik, 
{\sl Sov. Phys. JETP} {\bf 16 }, 220 (1963).

\bibitem{DER} 
N.N. Nikolaev and V.R. Zoller, {\sl JETP Lett.} {\bf 69}, 103 (1999).  

  
\bibitem{PION}
N.N. Nikolaev, J. Speth, and  V.R. Zoller, 
 {\em Phys. Lett.} B{\bf 473}, 157 (2000).
 
\bibitem{GamGam}
N.N. Nikolaev, J. Speth, and  V.R. Zoller, {\em J. EXP. Theor. Phys.}
{\bf  93},  957 (2001);
{\em Eur. Phys. J.} C{\bf 22}, 637 (2002).

\bibitem{CHARM2000}
N.N. Nikolaev and V.R. Zoller, {\em Phys. Lett.} B{\bf 509}, 283 (2001).

\bibitem{BEAUTY}
V.R. Zoller, 
 {\em Phys. Lett.} B{\bf 509},  69 (2001).

\bibitem{BFKL} 
L.N. Lipatov, {\em Sov. J. Nucl. Phys.} {\bf 23}, 338 (1976);
E.A. Kuraev, L.N. Lipatov, and V.S. Fadin, {\em Sov. JETP} {\bf 44}, 443 (1976);
 {\it ibid.} {\bf 45}, 199 (1977); Ya.Ya. Balitsky and L.N. Lipatov,
 {\em Sov. J. Nucl. Phys.} {\bf 28}, 882 (1978).

\bibitem{NZZ94}
N.N. Nikolaev, B.G. Zakharov, and V.R. Zoller,
{\em JETP} {\bf 105}, 1498 (1994).

\bibitem{NZDelta}
N.N. Nikolaev and B.G. Zakharov, {\sl Phys. Lett.} B{\bf 333}, 250 (1994);
  I.P. Ivanov and N.N. Nikolaev,
  {\em Phys.\ Rev.}  D{\bf 65} (2002) 054004.

\bibitem{NZHERA} 
N.N. Nikolaev, B.G. Zakharov, {\em Phys. Lett.}
 B{\bf 327}, 157  (1994).

\bibitem{NZcharm}
N.N. Nikolaev and  V.R. Zoller, {\em JETP Lett.} {\bf 69}, 187 (1999).

\bibitem{NZZ97}
N.N. Nikolaev, B.G. Zakharov, and V.R. Zoller, {\em  JETP Lett.} {\bf 66}, 138 (1997).

\bibitem{ELLIS}
J. Ellis, H. Kowalski, and D.A. Ross,  
 {\em Phys. Lett.} B{\bf 668}, 51  (2008).

\bibitem{H1ccbb} 
H1 Collab., A. Aktas et al, {\em Eur. Phys. J.}
C{\bf 45}, 23 (2006). 

\bibitem{H12008} 
H1 Collab. (preliminary),
 {\rm Measurement of  $F_2^{c\bar c}$ and $F_2^{b\bar b}$
using the H1 Vertex Detector at HERA.}
 Presented at 34th Int. Conf. on
 High Energy Physics, ICHEP2008, 30th July-5th August, 2008, Philadelphia.

\bibitem{H1FL2008} 
H1 Collab., F.D. Aaron et al, {\em Phys. Lett.}
B{\bf 665}, 139  (2008).

\bibitem{NZ91}
N.N. Nikolaev and B.G. Zakharov, {\em Z. Phys.} {\bf C49}, 607 (1991).

\bibitem{PISMA1} N.N. Nikolaev, B.G. Zakharov, and V.R. Zoller,
{\em  JETP\ Letters}\
 {\bf 59}, 8 (1994).

\bibitem{MEGGI}
M. D'Elia, A. Di Giacomo, and E. Meggiolaro,
{\em Phys. Rev.} D{\bf 67}, 114504 (2003).

\bibitem{TOLYA} 
A.V. Berezhnoy, V.V. Kiselev, and A.K. Likhoded,
{\em Phys. Atom. Nucl.} {\bf 63}, 1595 (2000).

\bibitem{White} C. D. White and R. S. Thorne {\em Phys. Rev.} D{\bf 74}, 
014002 (2006).

\bibitem{Baranov}
S.P. Baranov, {\em Phys.Rev.} D{\bf 76} 034021 (2007)
S.P. Baranov and  N.P. Zotov, {\em JETP Lett.} {\bf 86} 435 (2007);
S.P. Baranov,  A.V. Lipatov, and N.P. Zotov,  {\em Eur. Phys. J.} C{\bf 56}
371 (2008). 


\bibitem{ELBEAUTY} 
ZEUS Collab., J.Breitweg et al., 
 {\em Phys. Lett.} B{\bf 437}, 432 (1998); 
 H1 Collab., C.Adloff et al.,
 {\em Phys. Lett.} B{\bf 483}, 23 (2000). 

\bibitem{SCAN}
N.N. Nikolaev, {\em Comments on Nucl. Part. Phys.} {\bf 21}, 41 (1992);
B.Z. Kopeliovich, J. Nemchik, N.N. Nikolaev, and B.G. Zakharov, {\em Phys.Lett.}
B{\bf 309}, 179 (1993); ibid. B{\bf 324}, 469 (1994);
J. Nemchik, N.N. Nikolaev, and B.G. Zakharov, {\em Phys. Lett.}
B{\bf 341}, 228  (1994).

\bibitem{VMreview}
  I.P. Ivanov, N.N. Nikolaev and A.A. Savin,
  {\em Phys. Part. Nucl.}  {\bf 37} (2006) 1.

\bibitem{JETPVM}
J. Nemchik, N.N. Nikolaev, E. Predazzi,
  B.G. Zakharov, and V.R. Zoller, {\em JETP} {\bf 86}, 1054 (1998).

\bibitem{OP}
L.B. Okun and I.Ya. Pomeranchuk, {\sl Zh. Eksperim. i Teor. Fiz.}  {\bf 30}, 307 (1956);
{\sl Soviet Phys. JETP} {\bf 3}, 307 (1956).






































\end{thebibliography}
\end{document}